\documentclass[preprint,10pt,a4paper]{aastex}
\usepackage[colorlinks,urlcolor=blue,citecolor=blue,linkcolor=blue]{hyperref}
\usepackage{amsmath}
\usepackage{graphicx}
\usepackage{multicol}
\newcommand{\be}{\begin{equation}} 
\newcommand{\ee}{\end{equation}}

\newcommand{\ba}{\begin{eqnarray}}
\newcommand{\ea}{\end{eqnarray}}

\renewcommand{\v}{{\bf v}}

\newcommand\eg{\textit{e.g.,\ }}

\newcommand{\NS}{neutron star}
\newcommand{\NSs}{{neutron stars}}
\newcommand{\EM}{electromagnetic}

\newcommand{\ms}{magnetosphere}
\newcommand{\mss}{magnetospheres}

\begin{document}

\title{How else can we detect Fast Radio Bursts?}

\author{Maxim Lyutikov$^1$, Duncan R.~Lorimer$^2$}
\affil{$^1$Department of Physics and Astronomy, Purdue University, 
 525 Northwestern Avenue,
West Lafayette, IN
47907-2036, USA; lyutikov@purdue.edu\\
$^2$Department of Physics and Astronomy, West Virginia University,
 Morgantown, WV, 26506-6315
}
\begin{abstract} 
We discuss possible \EM\ signals accompanying Fast Radio Bursts (FRBs)
that are expected in the scenario where FRBs originate in
\NS\ magnetospheres. For models involving Crab-like giant pulses, no
appreciable contemporaneous emission is expected at other wavelengths.
Magnetar giant flares, driven by the reconfiguration of the \ms, however,
can produce both contemporaneous bursts at other wavelengths as well as
afterglow-like emission. We conclude
that the best chances are: (i) prompt short GRB-like emission; (ii) a
contemporaneous optical flash that can reach naked eye peak luminosity
(but only for a few milliseconds); (iii) a high energy afterglow
emission. Case (i) could be tested by coordinated radio and
high-energy experiments. Case (ii) could be seen in a coordinated  radio-optical surveys, \eg\ by the Palomar
Transient Factory  in a 60-second frame as a transient object of
$m=15-20$ magnitude with an expected optical detection rate of about
0.1~hr$^{-1}$, an order of magnitude higher than in radio. Shallow,
but large-area sky surveys such as ASAS-SN and EVRYSCOPE
could also detect prompt optical flashes from the more powerful 
Lorimer-burst clones. The 
best constraints on the optical-to-radio power for this kind of emission
could be provided by future observations with facilities like LSST. Case (iii)
might be seen in relatively rare cases that the relativistically
ejected magnetic blob is moving along the line of sight. 
\end{abstract}

\section{Introduction}

Fast radio bursts
\citep[FRBs;][]{2007Sci...318..777L,2012MNRAS.425L..71K,2013Sci...341...53T,2014ApJ...790..101S} are highly dispersed
millisecond-long radio events of unknown origin. Recently,
\cite{2016arXiv160302891L} argued that their properties are consistent
with non-catastrophic events in \NS\ \mss\ \citep[see also discussion by][]{2016arXiv160401799K}. In this framework, FRB progenitors  are
\NSs\ located at near-cosmological distances ($d \leq 100 $ Mpc) and
most of the dispersion comes from the local environment \citep{2015Natur.528..523M}, \eg\ a
new supernova shell or a dense young star cluster. The emission process
is either an analog of Crab giant pulses
\citep{2016MNRAS.457..232C,2016MNRAS.458L..19C,2016arXiv160302891L} or
a new, yet undetected, type of radio emission accompanying giant
flares in magnetars
\citep{lyutikovradiomagnetar,2010vaoa.conf..129P,2012MNRAS.425L..71K,2014MNRAS.442L...9L,2015ApJ...807..179P}. The
physical distinction is that giant pulses are rotationally-powered,
while magnetar flares are magnetically powered.

 \cite{2016arXiv160302891L} discussed the possible observational
 consequences of association of FRBs with pulsar giant pulses. The key
 point is the (proposed) scaling of FRB intensity with 
 spin-down power. If FRBs are (super)-giant pulses, no other
 contemporaneous \EM\ signals are expected.  The previous suggestion\footnote{This 
was based on
  the model of curvature emission origin of VHE photons which is
  currently disfavored for Crab \citep{2012ApJ...754...33L}.}
 by
 \cite{2007MNRAS.381.1190L} that giant pulses from Crab can be
 accompanied by GeV photons has not been  confirmed observationally
 \citep{2011ApJ...728..110B,2012ApJ...760...64M,2012ApJ...760..136A}.
  There
is a very mild correlation between giant pulses and X-ray emission
\citep{1995ApJ...453..433L}. This weak correlation is not likely to
be relevant for the near-cosmological FRBs.  Thus, if FRBs are
analogs of giant pulses, we do not expect any other contemporaneous
\EM\ signal.

As we discuss in this
{\it Letter}, prompt emission associated with magnetar giant flares allows 
for a wider
variety of contemporaneous \EM\ signals.
As for the origin of magnetar radio emission, a number of
authors \citep{lyutikovradiomagnetar,eichlerradiomagnetar,
  2010vaoa.conf..129P,2014MNRAS.442L...9L,2012MNRAS.425L..71K,2015ApJ...807..179P,2015arXiv151204503K}
discussed a possibility of prompt radio emission accompanying magnetar
flares. The predictions of \cite{lyutikovradiomagnetar} and
\cite{eichlerradiomagnetar}, which are based on somewhat different
models of radio emission, that magnetar radio spectra extend to
higher frequencies than in the rotationally-powered pulsars, have been
confirmed by observations \citep{2006Natur.442..892C}. Both models 
\citep{lyutikovradiomagnetar,eichlerradiomagnetar} used the scaling of
the frequency of the radio emission with the magnetospheric plasma
frequency, which in the case of magnetars can be much higher than in the
rotationally-powered pulsars \citep{tlk02}.
 
In this {\it Letter}, we discuss the expected properties of FRBs {\it if}
they are associated with magnetar giant flares. The main goal is to
highlight possible strategies for finding other \EM\ counterparts. The
association of FRBs with as yet undetected radio emission from magnetar flares is
even less certain theoretically than the case of giant pulses - no conclusion can be made from the first
principles. Also, transient radio emission of magnetars \citep{2006Natur.442..892C}  is
probably of different origin than in the rotationally-powered pulsars
\citep{lyutikovradiomagnetar,2016arXiv160302891L}.

\section{Magnetar giant flares: prompt emission at other wavelengths}

Using the observed radio fluxes of FRBs, and making educated guesses about
possible contemporaneous emission at other wavelengths we can estimate
the possibilities of detecting FRBs if they originate during magnetar
giant flares (and lower intensity bursts). Note that the Parkes
non-detection of radio emission during SGR~1806--20 giant flare
\citep{2016arXiv160202188T} provides arguments against the magnetar
association. Still, given that this observation was in the far
sidelobes of the Parkes beam, where it is hard to reliably measure
the true sensitivity of observations, we believe this remains a valid
possibility.

\subsection{Short GRB-like prompt emission}

The simplest case for contemporaneous \EM\ signal would be 
short GRB-like prompt emission from an FRB.  The prompt peak of emission of SGR
1806--20 flare could be seen to 40~Mpc \citep{palmer}.  Since
$\gamma$-ray monitors are wide-field (nearly full sky), there is no
need for special observations. Instead, all is needed is correlated time and
(wide-field) localization of any radio FRB with a short GRB.  Note,
that in this case the high-energy emission is not a typical short GRB,
coming presumably from the merger of two \NSs, but instead a similar looking
magnetar giant flare.  Also, the high energy range of $\sim 40$~Mpc
is somewhat lower than our basic estimate of FRB location within $\leq
100$~Mpc, so we expect that not every FRB is accompanied by a prompt
high-energy burst.

The physical requirements to produce a bright radio burst during a
magnetar flare are not constraining. Assuming that FRBs have  distances
$\sim 100$~Mpc, and taking the $F=1$~Jy peak flux as typical, 
as discussed by \cite{2016arXiv160302891L}, the instantaneous 
(isotropic-equivalent) luminosity
\be
L_{\rm FRB} =4\pi d^2  (\nu F_\nu)=  10^{40}  F_{\rm Jy} d_{\rm 100 Mpc}^2 {\rm erg \,\, s}^{-1},
\label{LFRB}
\ee 
where $F_\nu$ is the energy flux, as a function of frequency,
$\nu$, and we have assumed a flat spectrum.
In case of the magnetar flares, the peak $\gamma$-ray luminosity of
SGR~1806--20 flare was $10^{47}$ erg~s$^{-1}$. Thus, to produce an FRB,
Eq.~(\ref{LFRB}) implies that prompt radio efficiency $\eta_R$ needs to be only 
$\eta_R > 10^{-7}$.

Since the observed FRB pulse widths are often dominated by the propagation effects 
\citep[see, \eg\,][]{2016MNRAS.tmpL..49C}
more important is the total fluence ${\cal F}_{FRB} = F \tau$ for a
burst duration $\tau$. Considering a 5~ms pulse, this gives the total
emitted energy in radio $\sim 5 \times 10^{37}$~erg. This is much
smaller that the inferred total energy of $10^{46}$~ergs emitted in
$\gamma$-ray by the SGR 1806-20 flare. Below, when we scale various 
parameters with the radio properties of FRBs (\eg its duration) we 
always imply the scaling with the total fluence.

\subsection{Optical flashes}

If FRBs are related to explosive events like magnetar flares, we 
expect that, in addition to coherent prompt emission, the source also
produces non-coherent broad-band emission at other wavelengths (\eg in
optical and at high energies). We now discuss the energetics of 
possible counterparts.

For a possible optical prompt counterpart of an FRB with flux $F$, for
a flat spectrum source, the corresponding stellar magnitude 
\be
m= -2.5 \log_{10} \left(  \frac{F}{\rm 3631~Jy} \right).
\label{mFRB}
\ee
This expression implies that, if the peak energy flux in optical is
the same as in radio, the FRB will provide a very bright millisecond
flash of magnitude $m=8.9$! This estimate is very encouraging, since
typically (\eg for rotation powered pulsars) the
radio emission is a small fraction of total energetics and of emission at 
other wavelengths. For example, if in radio the efficiency is $\sim 100$ 
times smaller than in optical this would provide a naked eye optical flash 
lasting only a few milliseconds.

The Palomar Transient Factory \citep[PTF;][]{2009PASP..121.1395L} reaches
magnitude $m=20$ during a 60-s exposure.  A 5 millisecond flash
will give a fluence $\approx 10^4$ times smaller resulting in 
an image 10 magnitudes fainter. Since PTF is sensitive to a flash of
peak brightness of $m \sim 10$ - this will give the total 60-sec
exposure equivalent of $m=20$.  If optical power is $\eta_o$ times
larger than in radio, that can give an image of $m\approx 20- 2.5
\log_{10} \eta_o$.  It is reasonable to expect this to be $m\sim 15$.

\begin{figure}[!]
\includegraphics[width=0.99\linewidth]{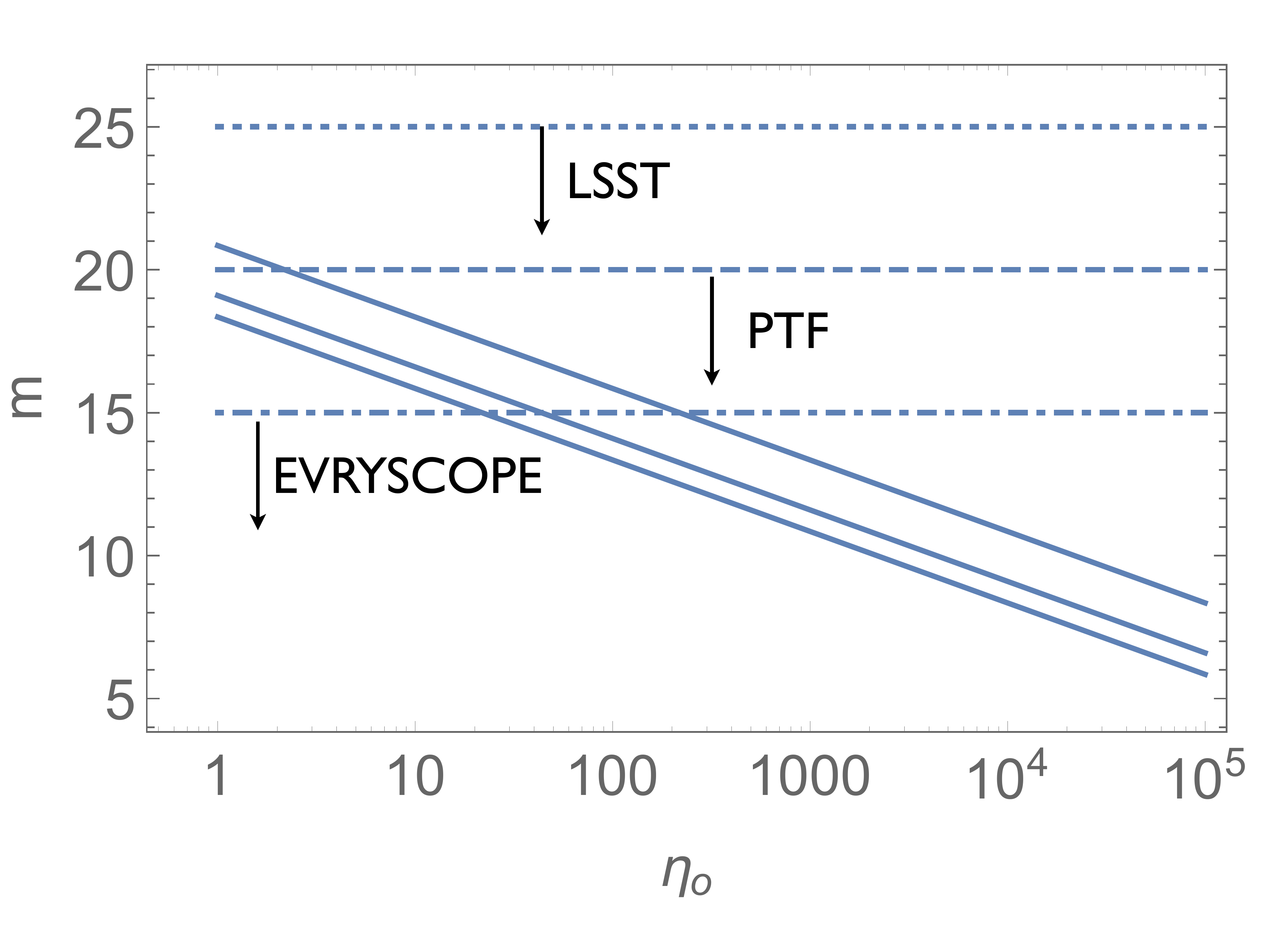}
\caption{Equivalent stellar magnitude of possible prompt optical emission
  from an FRB for a 60~s total exposure, as a function of $\eta_o$ -
  ratio of optical to radio energy fluxes for different FRB durations
  $\tau =1,\, 5\, 10$ msec (or, equivalently, for different total
  fluences). As examples of current and future optical transient survey
  facilities, 
  the expected sensitivities for PTF, LSST and EVRYSCOPE are plotted
  as the dashed horizontal lines.}
\label{m}
\end{figure}

More formally, using Eq.~(\ref{mFRB}), and scaling the optical flux to
the radio as $\eta_o= F_o/F_r \geq 1$, the expected magnitude 
\be
m= 20.8 - 2.5 \log_{10}  \left(\frac{\eta_o  \tau_{\rm ms} F_{\rm Jy}}{T_{60}} \right),
\label{mmm}
\ee
where $\tau_{\rm msec}$ is the pulse width in ms, $T_{60}$ is the exposure
time normalized to 60~s readouts and the peak flux density
$F_{\rm Jy}$ is in Janskys. This
dependence is shown for various assumed pulse widths in Fig.~\ref{m}.
Equation (\ref{mmm}) also highlights an important fact that shorter readout times are beneficial for the search of optical counterparts to FRBs. Other
current facilities that can be used to probe the parameter space shown in
Fig.~\ref{m} are PAN-STARRS \citep[$m \sim 24$;][]{2012ApJ...750...99T}, ATLAS
\citep[$m \sim 22$;][]{2015MNRAS.451.4238S}, ASAS-SN \citep[$m \sim 17$;][]{2016arXiv160400396H} and CRTS \citep[$m \sim $~19--20;][]{2009ApJ...696..870D}.

The forthcoming Large Synoptic Survey Telescope
\citep[LSST;][]{2008arXiv0805.2366I} is expected to fare even better.  It will also
have large field of view, almost 10 square degrees,
and will be able to reach magnitude
$m=25$ within two exposures of $\sim 15$~s each. So a 6~ms
flash has flux about 5000 times smaller than a 30~s exposure.
This would produce an image 9.3 magnitudes smaller.  Thus, LSST is
sensitive to a flash of peak brightness $m=15.2$, about 5 orders of
magnitude lower than PTF. In addition, a change in observing strategy - shorter readout times, down to one second -  might be used to increase effective sensitivity to
transients.

Much larger instantaneous fields of view are about to be explored to
shallower magnitude limits by ASAS-SN \citep{2016arXiv160400396H} and
EVRYSCOPE \citep{2015PASP..127..234L}. EVRYSCOPE
currently samples 8000~deg$^2$ fields of view every
two minutes. With a magnitude limit of $m = 16$, this instrument will
be less constraining in terms of $\eta_o$ (Fig.~\ref{m}) compared to 
PTF and LSST. However, as discussed below, it could provide excellent constraints
on optical emission from Lorimer-burst clones.

In conclusion, wide-field deep optical surveys with short exposure
time might be able to detect prompt optical emission from FRBs at a level of
$m=20$ magnitude. The likelihood of a detection would be even higher if 
the putative optical luminosity exceeds the
radio luminosity by several orders of magnitude. Alternatively, null results
from PTF and LSST in future will provide important constraints on prompt
optical emission.

\subsection {Rates for contemporaneous detections}

The all-sky FRB rate of $4.4^{+5.2}_{-3.1} \times 10^3$ per day
\citep{2016MNRAS.455.2207R} implies approximately 0.2--1.6 FRB per 8 square
degrees per day (equivalent to PTF or LSST fields of view), 
or 8--80 FRB per 8000 square degrees per hour (equivalent to EVERYSCOPE).  
PTF and LSST should, therefore, have an
FRB in their field of view every $\sim 12$--90 hours. Since LSST has a very
similar field of view this estimate also applies. The average rate
of FRB detections at Parkes, based on the survey of
\cite{2015arXiv151107746C} which detected nine FRBs in 2500~hr of
observing, is approximately 280 hours per FRB.

The estimate in Equation
(\ref{mmm}) for the brightness of an optical counterpart
can be scaled to other survey telescopes. For example, the BlackGEM
telescope \citep{2015ASPC..496..254B} will consist of fifteen
telescopes covering total area of 40 square degrees down to $m\sim 20$
in one minute exposure. With the sensitivity similar to PTF, but four
times larger area the expected detection rate will be one per few
hours. Importantly, using continuous read-out BlackGEM can go to much
shorter exposure times, down to few seconds -- a transient flash will
look brighter for shorter exposure times.

These estimates show that contemporaneous radio-optical detection
will be determined mostly by the (relatively small) field of view of
radio telescopes. While LOFAR \citep{2013A&A...556A...2V},
with its wide-field coverage might prove effective, \eg in
post-processing analysis of the optically identified flashes,
free-free absorption at low frequencies might be important
\citep{2016arXiv160302891L,rl16}.
The best prospects for contemporaneous monitoring will most likely
come from CHIME \citep{2014SPIE.9145E..22B} or UTMOST \citep{2016MNRAS.458..718C}, 
where the predicted rates could be up to hundreds of FRBs per day
\citep{rl16}.

For EVRYSCOPE, the high event rate but shallow sensitivity most
likely makes it suitable to bursts from the brightest FRBs. Taking 
the all-sky event rate for such bursts
estimated to be around 250 per day \citep{2007Sci...318..777L}, we find that of
order 2 similar bursts would be seen in the EVRYSCOPE field each hour.
Inserting $F_{\rm Jy}=30$ and $\tau=15$~ms into Eq.~\ref{m}, 
we find $m=15$ for a 2~min exposure
if $\eta_o=1$. This would be eminently detectable by EVERYSCOPE for which
the limiting magnitude is 16.4 in 2~min \citep{2015PASP..127..234L}.

\subsection{Radio  and high energy GRB-like afterglows}
 
The 2004 flare from SGR 1806--20 produced a total of $4 \times 10^{43}$~ergs
radio afterglow with peak flux 50~mJy \citep{Gaensler1806} from a
distance of few kpc. Given that the total energetics of the
SGR 1806--20 flare was $10^{46}$ ergs \citep{palmer}, 
the likely radio afterglow efficiency is $4 \times 10^{-3}$.
Since FRBs are at $d>$~few~Mpc, the expected peak flux from a similar
radio afterglow would be $< 50$~nJy and therefore would not be detectable
by current telescopes. 
At higher energies, no afterglows were observed for SGR 1806--20
\citep{Gaensler1806}. Since FRBs are
$\sim 10^{3}$ times further away, we do not expect to see appreciable
high-energy afterglows.
 
On the other hand, there is a possible caveat in the above argument
against detectability of afterglow emission.
\cite{2006MNRAS.367.1594L} suggested that though the initial giant
flare spike is quasi-isotropic, the ejected relativistically moving
magnetic blob (an analog of Solar coronal mass ejections, CMEs) has
been collimated into an opening angle $\sim 0.1$~radians.  In the case of SGR
1806--20, the motion of the blob was directed away from the observer, so
that its emission was de-boosted. If the relativistic CME is
directed along the line of sight, we can indeed expect an afterglow
which emission is boosted towards an observer lasting $\tau_{\rm ag}
\sim$ few weeks. The observed isotropic equivalent luminosity of such
afterglow can reach isotropic equivalent luminosity $L_{\rm ag} \sim
E_{\rm ej} /( \tau_{\rm ag} \Delta \Omega) \sim 10^{42} $ erg s$^{-1}$, where
$ E_{\rm ej} \sim 10^{46} $ erg is the total energy contained by the CME,
$ \tau_{\rm ag} \sim 10$ days is the duration of the afterglow and $
\Delta \Omega \sim 10^{-2}$~sr is the solid angle of the boosted emission
by the relativistically moving CME. For a source located at $\sim 100$
Mpc the corresponding flux at the Earth is relatively high, $ \sim 10^{-14}
$~erg~cm$^{-2}$~s$^{-1}$.  Such transient sources could be 
detected by existing instruments (\eg\ XMM sensitivity reaches $ \sim
10^{-16} $ erg cm$^{-2}$ s$^{-1}$).  The key drawback of this
possibility is that only a small number of magnetar giant flares,
$\sim \Delta \Omega \sim 10^{-2}$~sr, is expected to produce CMEs moving
along the line of sight.

We note in this respect that the radio afterglow for FRB~150413 claimed by
\cite{2016Natur.530..453K}, which was recently argued to be due to AGN
variability \citep{2016ApJ...821L..22W,2016arXiv160304421V},
 generally agrees with these estimates. For
example, the afterglow reported by \cite{2016Natur.530..453K} needs
about $10^{45}$~ergs emitted in radio. This is different by about two orders
of magnitude from the 2004 flare from SGR 1806$-$20. This increased
afterglow luminosity can be due to the mildly relativistic ejection of
the CME toward the observer. 
 
\section{Summary}
 
In summary, we have discussed several possible strategies for
observing possible \EM\ counterparts of FRBs. If FRBs are related to
rotationally-powered giant pulses from newly born ``super''-Crab
pulsars, we do not expect any other \EM\ signal.  If FRBs are related
to magnetar giant flares we can expect (i) to detect the prompt high
energy flare; (ii) contemporaneous optical flash that in a $\sim $ 60
seconds exposure can reach equivalent magnitudes of $m\sim 15-20$. In
fact, the rate of optical flashes to be seen by PTF and LSST are
expected to be higher than the rate of FRBs detected by high frequency
searcher. Finally, afterglows can be detected in rare circumstances
than the magnetic blob ejected during magnetospheric reconfiguration
is moving relativistically towards the observer.

We would like to thank Edo Berger, Paul Groot, Sergey Popov and Matt Wiesner for discussions. This work was supported by  NASA grant NNX12AF92G and NSF grants AST-1306672 and AST-1516958.

\bibliographystyle{apj} 

\end{document}